# Cosmic muon flux at shallow depths underground.

*L.N.Bogdanova, M.G.Gavrilov, V.N. Kornoukhov, A.S.Starostin,*

*Institute for Theoretical and Experimental Physics, B.Cheremushkinskaya ul, 25, Moscow 117218, Russia*

**Abstract**

We consider the cosmic muon background for the installations located at shallow depths. We suggest a relatively simple formula for the sea-level muon spectrum, which allows calculate dependencies of the vertical muon intensity and integral muon flux density on overburden. Muon flux dependency on the zenith angle at overburden of 10 to 100 meters of standard rock shows that muon angular distribution practically does not change in this interval. We present muon angular distributions for three typical apparatus locations in measurements on the surface and at shallow depths. It is shown that for such installations the active shielding "umbrella" should overlap a zenith angle of θ ~ 80° to remove the cosmic muon background.

**1. Introduction**

Intensity of the muon component of cosmic rays (CR) at relatively small depth underground (from tens to several hundred meters of water equivalent (m w.e.)) is of practical interest for many fundamental studies and applied research based on the low-background measurements. These might be reactor experiments on neutrino oscillations [1-3], searches of the neutrino magnetic moment at reactors [4,5] or with the artificial antineutrino source [6], studies associated with measurements of low activities in materials. CR background at the sea level consists of three main components:

- Soft (electrons and γ-quanta of electromagnetic showers);
- Hadronic (hadrons of nuclear showers);
- Hard ($\mu^{+/-}$).

Partition of sea-level CR flux is shown in Table 1.

Table 1. Percentage of sea-level CR flux components [7].

| Total flux | Muons | Secondary neutrons | Electrons | Protons, pions |
|---|---|---|---|---|
| $3 \cdot 10^{-2}$ cm$^{-2}$s$^{-1}$ | 63% | 21% | 15% | < 1% |

For practically complete suppression of the soft CR component a lead layer of ~15 cm thickness is enough. The charged CR component can be registered very efficiently; hence this background can be easily removed by using anti-coincidences (AC). Complications with suppression of the charged CR component come up when installation surface is large: $\geq 100$ m$^2$, when numerous counts in AC system result in a noticeable loss of useful events (experiments on neutrino oscillations). A more serious problem is a removal of the neutron background. Secondary neutrons are generated by hadronic CR component in the atmosphere. At sea level their flux density is $5 \cdot 10^{-3}$ n·cm$^{-2}$·s$^{-1}$ and decreases exponentially with depth ($\lambda \sim 165 \div 200$ g·cm$^{-2}$, depending on the medium properties). To suppress the secondary neutron flux by a factor of 100, a shielding thickness of about ~1000 g·cm$^{-2}$ is required, or ~ 4 m of a standard rock (SR)[1].

Tertiary neutrons emerge in the electromagnetic processes when high-energy muons pass through matter and in the capture of slow muons by nuclei. Since large penetrating ability is specific for muons, intensity of accompanying tertiary neutrons decreases with depth rather slowly. This makes noticeable difficulties for experimentalists, especially if neutrons are produced in the passive shielding of the detector [8]. In slow muon capture by nuclei, hundreds of microseconds can pass from the moment of muon capture to Bohr orbit till its absorption by a nucleus followed by neutron emission. Such background events are removed by AC system by working out a forbidding signal of matching duration. To make anti-coincidences effective and do not pass over useful events, it is necessary to have reliable information on the intensity and angular distribution of the muon flux (MF) in the setup location.

For today there are plenty of works devoted to MF calculation. However, most of them were performed for specific experimental installations located at depths of several thousand m w. e.. Besides, these calculations are exclusive, since they take into account the map of overburden and the properties of the local medium specific for the given experiment, and more important, disregard a soft component of the muon flux, which is inessential for large depth. All this makes their direct usage for determining MF at shallow depths difficult.

**2. Calculation of muon flux at small depth.**

---

[1] A standard rock is a ground with the following parameters: density $\rho = 2.65$ g/cm$^3$, $<Z/A> = 0.5$; $<Z^2/A> = 5.5$.

Ground-based experimental installations and those located at shallow depth appear in a variety of conditions with respect to absorbing substance. Low-background spectrometers can be placed in open shafts [3], be located under relatively small overburden [8], or on lower floors of the buildings [9], where concrete constructions and equipment of upper floors play the role of passive shielding. For example, power reactors, where experiments on the neutrino magnetic moment measurement are carried out, provide the effective thickness of the passive shielding of (30÷60) m w. e.. In all above cases for the calculation of the integral MF, J [cm$^{-2}$·s$^{-1}$], it is necessary to know the space distribution of MF density in the setup location. A common characteristics of the MF variation with depth is its vertical intensity $I_\perp$[cm$^{-2}$·s$^{-1}$·sr$^{-1}$]). At sea level the vertical flux intensity of muons of energy above 1 GeV is $I_\perp \sim$ 70 m$^{-2}$·sr$^{-1}$·s$^{-1}$ [10].

The intensity of muons underground can be calculated from the muon spectra at sea level and the rate of their energy loss. In principle, sea-level muon spectra in various angular intervals can be taken from the direct measurements. However, it is known that numerous sea-level measurements for the vertical muon flux $I_\perp$ and for near-horizontal flux are in rather poor agreement with one another [7,11]. In fact the experimental accuracy varies from 7 % at 10 GeV to 17 % at 1 TeV [11]. Hence, it would be practical to use an analytical formula reasonably describing the totality of the data, for the muon background calculation.

There exist several analytical (fitting and semi-empirical) expressions for the muon angular and energy distribution at sea level in the literature [12-20]. Results of various theoretical calculations of the sea-level differential muon spectra $I_\perp$ in the energy range 10÷10$^5$ GeV are usually tabulated. Corresponding fitting expressions are not universal over the whole energy interval and, in practice, consist of several segments that describe MF satisfactorily in limited energy intervals.

For our calculation of the muon flux depth dependency we used the following approximation of the sea-level muon spectra:

$$\frac{dI(\theta,\varphi,p)}{dp} = \frac{18}{p\cos\theta + 145}(p + 2.7\sec\theta)^{-2.7}\frac{p+5}{p+5\sec\theta}. \qquad (1)$$

Muon momentum p is measured in GeV/c, $1 \leq p \leq 10^5$ GeV/c. The structure of formula (1) is similar to that from paper [12], but by parameter choice we better describe the experimental spectrum of the vertical MF density at p < 10$^3$ GeV/c and spectra in the zenith angle range from 0° to 85° [21]. As for discrepancies noticeable at larger zenith angles, according to our analysis, they prove to be inessential for the results. Approximating formula (1), being rather simple, does not contradict to results of theoretical calculations of the vertical muon flux (Fig.1).

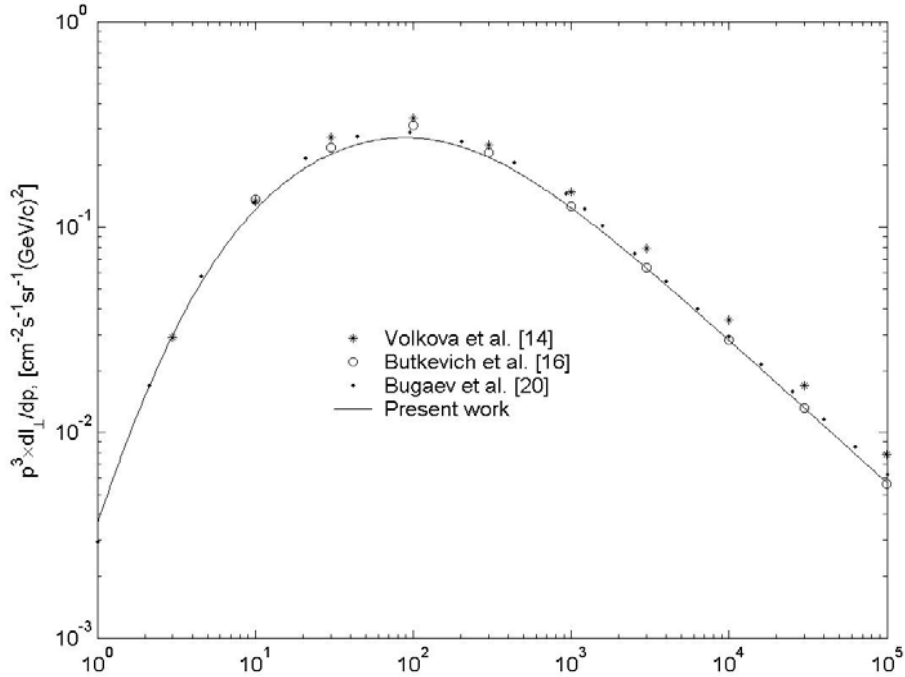

Figure 1. Spectrum of the vertical muon intensity at sea level calculated in the present paper and in papers by Volkova *et al* [14], Budkevich *et al* [16], Bugaev *et al* [20].

Using formula (1) and tables of muon range versus energy [22], we calculated dependencies of the vertical intensity and integral MF on the standard rock overburden. The integral muon flux is the number of muons passing through the hemisphere of unit surface per time unit ($m^{-2} \cdot s^{-1}$). It is obtained by integrating of $I(\theta,\varphi)$ over the solid angle. Muon flux distribution over the zenith angle $\theta$ is taken into account, dependency on azimuth angle $\varphi$ being uniform. Results of our calculations are shown in Fig.2. Calculations were performed in the energy interval $1$ GeV$<E<10^5$GeV. Curvature of the earth surface was not taken into account.

### 3. Muon flux angular distributions at shallow depth.

Zenith angle $\theta$-dependency at different depths is of interest by itself. As it is seen from results for 20, 40 and 100 m of SR depth presented in Fig.3, these distributions turn out to be very similar to the distribution at the surface, contrary to the frequent expectation that they should become narrower, since the absorption length increases with angle. However, the average energy of the muon spectrum (and the muon range) increases with angle [21,22]. Thus, when estimating the MF at depths from 0 to 100 m of SR, one can use the angular distribution function $I(\theta) \approx I_\perp \cdot \cos^2(\theta)$, (Fig.3). It is seen from Fig.3 that it satisfactorily describes all angular distributions within this interval of overburden. The same conclusion can be drawn from the results presented in Fig.2. The ratio of vertical intensity and integral MF is practically constant for depths down to 100 m of SR, which indicates the unvarying character of connection between them.

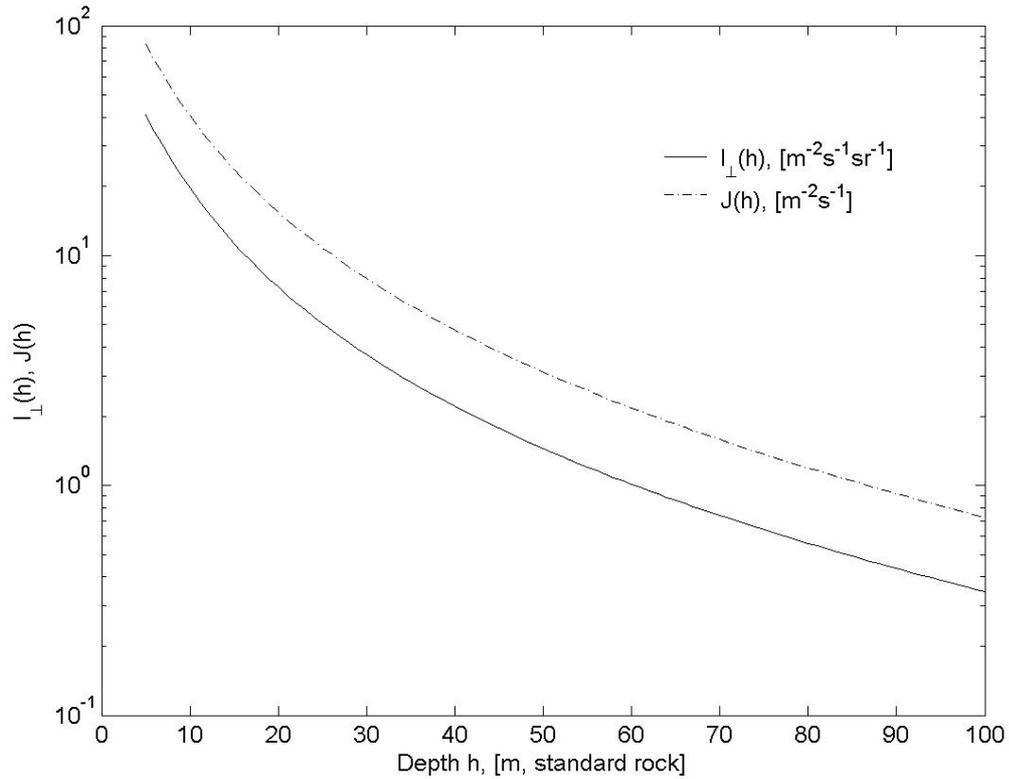

Figure 2. Vertical intensity and integral muon flux versus the standard rock overburden.

This is not the case when muon angular distributions under solid overburden are compared with those in the open shaft and under a reactor. To emphasize distinctions in angular distributions, locations of the installations were chosen to be adequate to real conditions and more or less equivalent by overburden. The overburden of 20 m SR was taken in the first case. The shaft depth was chosen to be 20m, its diameter was 6 m. Equipment and constructions above the installation are comparable to passive shielding of 3 m SR. In case of spectrometer under the reactor, equivalent thickness of passive shielding at zero zenith angle was also taken to be 20 m SR. Since it is rather difficult to take into account the non-uniform distribution of the absorbing medium in the reactor, we performed a model calculation of angular distributions for an object located under the water-filled cylinder of 50 m height and 30 m diameter. Our calculations for these configurations are shown in Fig.4.

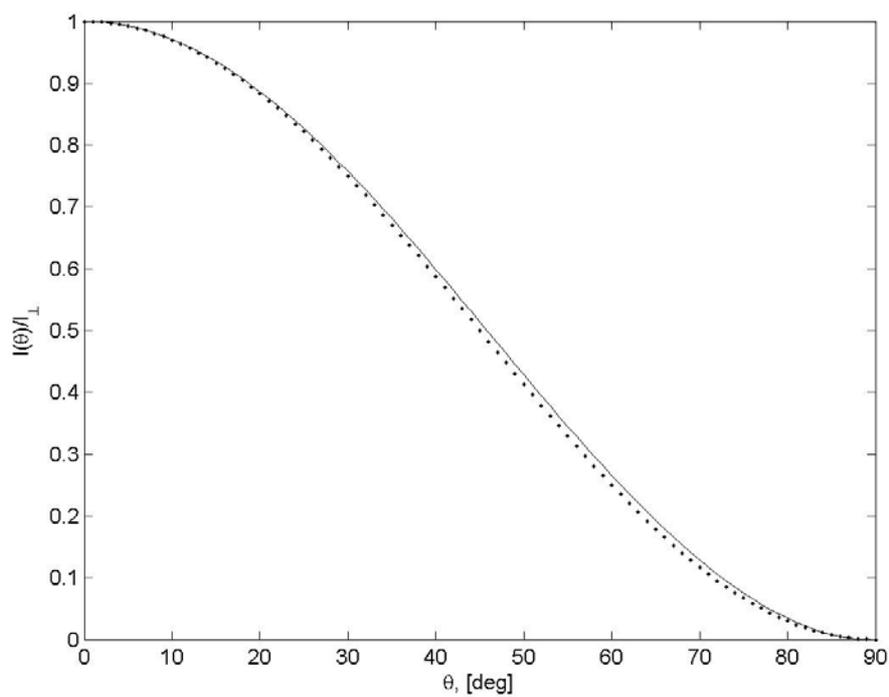

Figure 3. Normalized muon angular distributions $I(\theta)/I_\perp$ at depths from 10 to 100 m of SR –solid curve. Dotted line – distribution $I(\theta)/I_\perp = \cos^2(\theta)$.

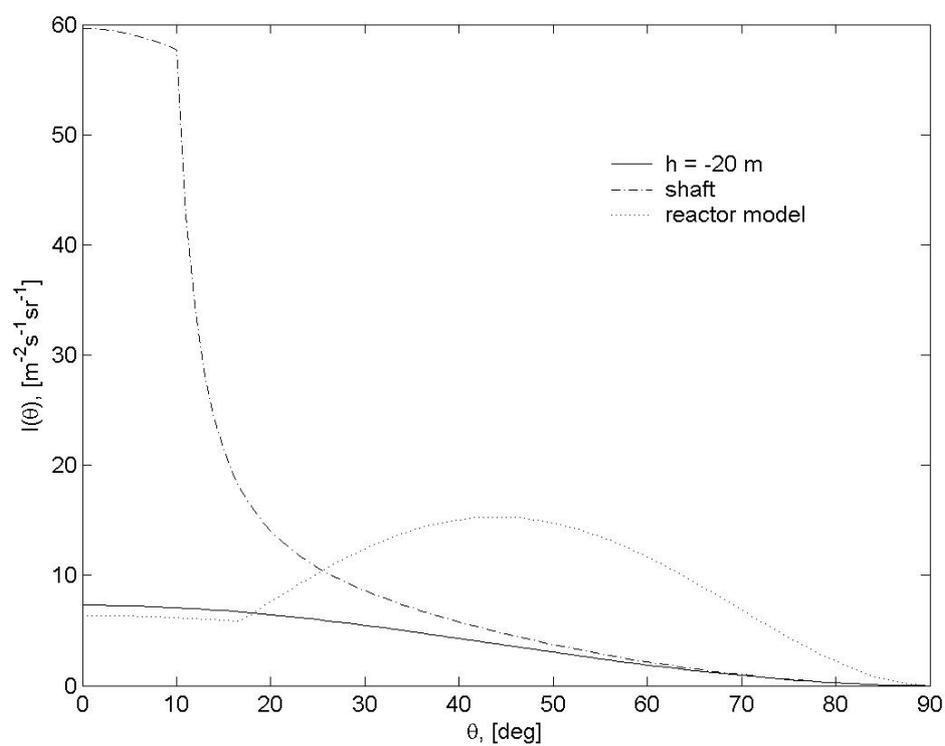

Figure 4. MF dependency on azimuth angle $\theta$ for installations located underground, in the shaft and under the reactor.

While designing the active shielding it is important to know the total number of muons that fall onto the installation at definite zenith angle from all azimuth directions $I_\Sigma = I(\theta)2\pi \cdot \sin(\theta)$. Calculations of $I_\Sigma$ for the standard rock overburden, shaft and reactor are shown in Fig.5. For all cases minimum of the distribution corresponds to zero zenith angles, while positions of the maxima differ essentially.

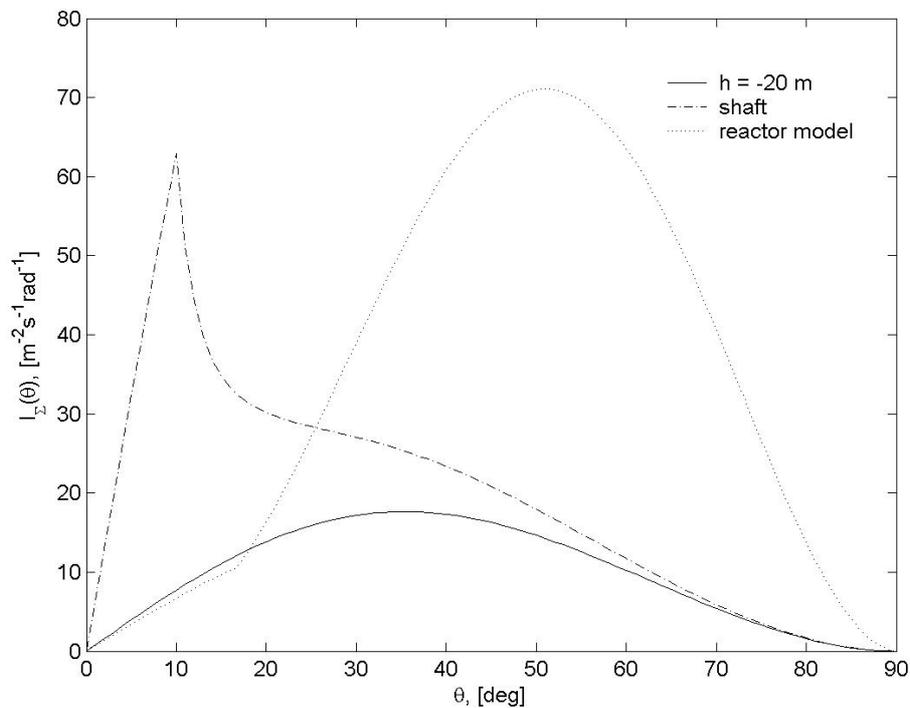

Figure 5. Muon flux $I_\Sigma$ dependency on angle $\theta$ for objects located underground, in a shaft and under the reactor.

We can obtain very useful distributions by integrating distributions in Fig.5 over $\theta$ in the interval $0 \leq \theta \leq \theta_{AS}$ and normalizing them to the integral MF. They are shown in Fig.6 and demonstrate the fraction of the integral muon flux that is captured by the active shielding (AS) "umbrella", depending on the covered angle. Often the AS covers the installation from above, $\theta_{AS} \approx 30° \div 40°$. It follows from the results shown in Figs.5 and 6 that such a shielding, covering zenith angles of ~30°÷40°, in most cases does not solve the problem of the cosmic background removal.

## 4. Conclusions.

Our calculations of the cosmic ray muon component at overburden from tens to hundreds meters of water equivalent might be of relevance for designing installations for fundamental studies and applied research based on low-background measurements. We make an effort to take accurately into account the soft part of the sea-level muon spectrum in our calculation of the depth dependency of the vertical and integral muon flux. This was not important in previous studies addressed mainly to deep overburden. In addition to the integral muon flux intensity, its angular distribution is also of practical significance for the low-background measurements.

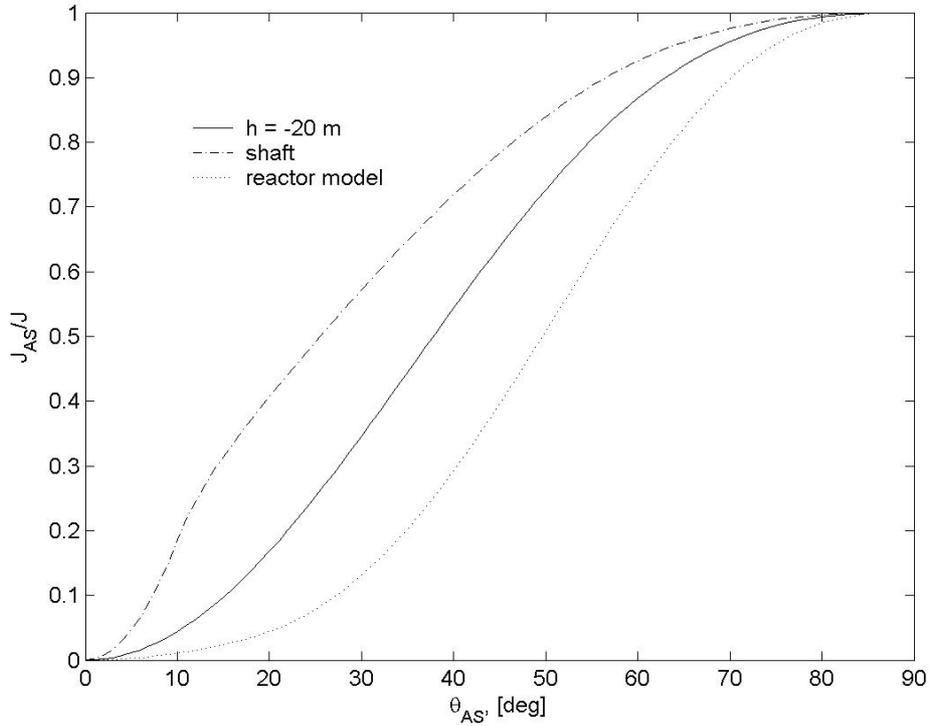

Figure 6. Dependency of the MF fraction captured by the active shield overlapping zenith angle $\theta_{AS}$ for objects located underground, in the open shaft and under the reactor.

Our calculations demonstrate that the muon flux angular distributions remain practically unchanged at overburden, at least from 10 to 100 meters of SR. Our results give the idea of the integral muon flux and its angular distribution for a diversity of experimental installations, located in open shafts, under buildings and under small overburden. It is shown that for such installations the active shielding "umbrella" should overlap a zenith angle of $\theta \sim 80°$ to remove the cosmic muon background.

We believe that suggested formula and presented charts can help obtain numerical estimations before more detailed calculation of the muon flux for a specific location and properly design an active shielding for an installation.

**References**

1. L.Mikaelyan, Nucl.Phys.B (Proc.Suppl.) **91** (2001) 120; V.Martemyanov, L.Mikaelyan, V.Sinev, et al., Yad.Fiz. **66** (2003) 1982; [Phys.At.Nucl. **66** (2003) 1934].